\documentclass[twoside,10pt]{article}

\usepackage{titlesec}
\usepackage{latexsym}
\usepackage{theorem}
\usepackage{fancyhdr}
\usepackage{amssymb}
\usepackage{amsmath}

\renewenvironment{abstract}[1]{\begin{minipage}[t]{\textwidth}
\mbox{\Large #1} \hspace{-0.5em}}{\end{minipage}}

\newcommand{\emphbf}[1]{\textbf{#1}}
\newcommand{\nn}{\nonumber}

\newcommand{\beq}{\begin{equation}}
\newcommand{\eeq}{\end{equation}}
\newcommand{\beqarr}{\begin{eqnarray}\vspace{1em}}
\newcommand{\eeqarr}{\vspace{1em}\end{eqnarray}}

\newcommand{\bdefinition}{\begin{definition}}
\newcommand{\edefinition}{\end{definition}} 
\newcommand{\btheo}{\begin{theorem}\vspace{1em}}
\newcommand{\etheo}{\vspace{1em}\end{theorem}}

\newcommand{\laction}{\triangleright}

\newcommand{\tensor}{\otimes}
\newcommand{\id}{\textrm{id}}
\newcommand{\coproduct}{\Delta}
\newcommand{\counit}{\epsilon}
\newcommand{\one}{\mathbf{1}}
\newcommand{\antipode}{\textrm{S}}

\theoremstyle{change}

\theoremheaderfont{\scshape}

\newtheorem{definition}{Definition}[section]
\newtheorem{theorem}[definition]{Theorem}

\titlespacing{\section}{0em}{0em}{2em}
\titlespacing{\subsection}{0em}{2em}{2em}

\titleformat{\section}
  {\sc \large}
  {\thesection}
  {1em}
  {}
\titleformat{\subsection}
  {\sc}
  {\thesubsection}
  {1em}
  {}

\begin{document}


\title{\scshape Construction of $\theta$-Poincar\'e Algebras and their Invariants on $\mathcal{M}_\theta$}
\author{\renewcommand{\thefootnote}{\arabic{footnote}}\scshape Florian Koch \footnotemark[1] $\;$, 
Efrossini Tsouchnika \footnotemark[2] \\[4em]
\rule{20mm}{0.2mm} \\[1em]
\small \scshape ${}^{1,2}$Sektion Physik der Universit\"at M\"unchen,\\[-0.5em]
\small \scshape Theresienstra{\ss}e 37, 80333 M\"unchen, Germany\\[1.5em]
\rule{20mm}{0.2mm} \\[2em]}
\date{}

\footnotetext[1]{koch@theorie.physik.uni-muenchen.de}
\footnotetext[2]{frosso@theorie.physik.uni-muenchen.de}

\maketitle


\begin{abstract}{I}n the present paper we construct deformations of the Poincar\'e algebra 
as representations on a noncommutative spacetime with canonical commutation relations. These 
deformations are obtained by solving a set of conditions by an appropriate ansatz 
for the deformed Lorentz generator. They turn out to be Hopf algebras of quantum universal 
enveloping algebra type with nontrivial antipodes. In order to present a notion of 
$\theta$-deformed Minkowski space $\mathcal{M}_\theta$, we introduce Casimir operators 
and spacetime invariants for all deformations obtained.
\end{abstract}

\setcounter{footnote}{1}
\thispagestyle{empty}



\newpage

\section{Introduction}
An important issue of high energy physics is the unification of all physical interactions
into a single renormalized quantum field theory. Most of the various approaches to this aim 
share the idea that both topics, unification and renormalization, should be addressed 
simultaneously by the introduction of a natural unit of length. The concept of a natural unit of 
length is almost as old as quantum field theory itself, since the cut-offs that ensure the
finiteness of integrals over momentum space in ordinary perturbation theory actually correspond to
nonlocal interactions that general relativity already suggests by the Planckian length

$$ \lambda_p = (\frac{G \hbar}{c^3})^{\frac{1}{2}} \simeq 1.6 \times 10^{33} \textrm{cm}. $$

This fundamental unit of length marks the scale of energies and distances at which nonlocality
of interactions has to appear and a notion of continuous spacetime becomes meaningless. Such nonlocal 
interactions have been introduced in various ways: the approaches to this matter
range from rather mathematical concepts to fundamental physical motivations - depending on whether 
the aspect of renormalization or that of general relativity is emphasized.

Apart from string theory the oldest and rather abstract path to the topic is that of noncommutative
geometry. From the physical point of view this corresponds to a generalization of the quantization
scheme by replacing the real valued coordinates of configuration space by a noncommutative algebra of
hermitean operators. One of the earliest and most prominent examples is Snyder's work from the 1940s 
where he showed that, under the requirement of Lorentz covariant spacetime spectra, any introduction 
of a finite natural unit of length necessarily leads to a noncommutative algebra of 
coordinates \cite{snyder}. Although Snyder's model exhibits intriguing properties that could lead to 
renormalization of self-energies and vacuum polarizations, it also features the main trouble 
noncommutative geometry has to deal with, namely the breaking of spacetime 
symmetries\footnote{Snyder's construction leads to a de Sitter momentum space that brakes translational 
invariance. This problem was solved by Yang \cite{yang}. Further field theoretical aspects 
were studied by Gol'fand \cite{golfand} later.}.

In the 1980s and early 1990s noncommutative spaces and their symmetries were investigated more 
systematically in the context of quantum groups which arose from the work of L.D. Faddeev on the 
inverse scattering method \cite{faddeev}. The first objects studied in quantum groups were 
deformed Lie algebras and groups such as $U_q(sl_2)$ of P.P. Kulish and N.Y. Reshetikhin \cite{kulish} 
or compact quantum matrix groups such as $SU_q(2)$ of S.L. Woronowicz \cite{woronowicz}. 
These quantum groups were identified to be Hopf algebras as E.K.Sklyanin showed for example for 
$U_q(sl_2)$ in \cite{sklyanin}. Moreover V.G. Drinfeld and M. Jimbo found a whole class of one 
parameter deformations of semi-simple Lie algebras \cite{drinfeldjimbo} being Hopf algebras
of quantum universal enveloping algebra type. The study of representations always kept 
the contact to physical aspects. At the beginning of 1990s q-deformations of the Lorentz 
and Poincar\'e algebra on a q-Minkowski space \cite{qminkowski} were obtained.

Despite their elegance and their mathematically rigorous construction these noncommutative spaces turn
out to be far too complicated to construct field theories on them with a reasonable amount of effort. 
This is mainly due to the fact that the commutation relations of the corresponding quantum spaces 
are fully quadratic in the coordinates. This makes it impossible to define Moyal-Weyl star products 
in terms of exponential expressions. Prominent exception is the $\kappa$-Minkowski space \cite{lukierski} 
that allows a study of field theoretic aspects as for example in \cite{munichkappa}.

Parallel to the development of quantum groups, string theory blazed its trail to be the first serious
attempt to unification and renormalization. In the last years open strings with homogenous magnetic
background field \cite{stringbackground} gave rise to so called brane world scenarios where
the effective field theories live on noncommutative spaces with canonical commutation
relations\footnote{A similar result was received at the beginning of the 1970s where the effective
theory of charged particles in a homogenous electric field lead to the same noncommutative space
\cite{schrader}.}. Seiberg-Witten map \cite{seibergwitten} and deformation quantization
\cite{deformation} opened the doorway to gauge theories on noncommutative spaces that even lead to
noncommutative versions of the standard model of particles and the grand unified theory
\cite{munichtheta}. The main drawback in this latest approach is the absence of a scheme of 
quantization and of spacetime symmetries other than translation invariance. Note that noncommutative 
spaces with canonical commutation relations were also obtained by introducing the nonlocality by 
general relativistic arguments, where the involved constructions  become covariant under Lorentz 
symmetries by imposing additional quantum conditions on the antisymmetric constant tensor 
\cite{fredenhagen}. However, since modern approaches to quantum gravity, as loop quantum 
gravity, suggest that not only the configuration space but also the symmetry algebra should 
be deformed \cite{smolin}, we consider an alternative way via quantum groups in the present paper.

We present here an attempt to join the area of quantum groups with that of field theories on 
$\theta$-spacetime structures. We derive $\theta$-Poincar\'e algebras as representations on the
noncommutative spacetime algebra with canonical commutation relations. These deformations turn out to
be Hopf algebras of quantum universal enveloping algebra type. 

As a generalization of the present results we also sketch how one could possibly obtain 
deformed symmetry algebras to any \emph{given} noncommutative space. In the traditional approach 
of quantum groups the algebraic properties of quantum spaces are determined by the deformation applied 
to the symmetry algebra. Here we fairly follow the opposite way - however this generalization is 
still work in progress. 

In the first part of this paper we collect all requirements that any deformation of a symmetry
algebra to any given noncommutative space has to obey. For $\theta$ - Poincar\'e algebras we restrict 
ourselves to the case of quantizations that are linear in the deformation parameters and show how 
deformations arise by the choice of an appropriate generating ansatz for the deformed Lorentz operator. 
We find continously many solutions that all turn out to be Hopf algebras. One of the solutions we present 
here was also found independently by alternative considerations by M. Dimitrijevi\'c et al. 
in the attempt to study concepts of 
derivatives on deformed coordinate spaces \cite{wess}. Furthermore the same solution was obtained 
by the authors of \cite{chaichian} using a suitable Drinfeld-twist. 

Finally we work out explicit expressions for the Pauli-Lubanski vector and the configuration space 
invariant for all solutions. Thus we obtain a notion of $\theta$-Minkowski space $\mathcal{M}_\theta$.

Our considerations incorporate the following notations and conventions. In general the field and quantum 
group theoretic aspects of our considerations orient themselves to the references \cite{fieldtext}
and \cite{quantumtext}. We use latin and greek letters for indices of space and spacetime coordinates 
respectively
\beqarr
   i,j,k,\ldots &\in& \{1, 2, 3\} \nn \\
   \mu,\nu, \ldots &\in&  \{0, 1, 2, 3\}. \nn
\eeqarr
Our presentation is independent of any specific choice of the signature for the metric tensor 
$\eta^{\mu\nu}$ in commutative Minkowski space $\mathcal{M}$.

\vspace{2em}

\section{The Poincar\'e Algebra and its $\theta$ - Deformations $U^\lambda_\theta(\mathfrak{p})$}

In this section we derive $\theta$-deformations of the Poincar\'e algebra $\mathfrak{p}$ as 
representations on noncommutative spacetime algebras $\mathfrak{X}_\theta$ with canonical 
commutation relations
\beq
        [x^\mu,x^\nu ] = i\theta^{\mu\nu}. \nn
\eeq
We find continuously many solutions of quantum universal enveloping algebra type, 
$U^\lambda_\theta(\mathfrak{p})$, that are parametrized by real 
parameters $\lambda = (\lambda_1, \lambda_2)$.

The section contains three parts. In the first subsection we collect a set of three conditions 
that any deformation of the type $U_\theta(\mathfrak{p})$ has to satisfy. In the second part 
we present an ansatz for the operators $M^{\mu\nu}$ of the deformed Lorentz algebra that generates 
the desired solutions $U^\lambda_\theta(\mathfrak{p})$. Finally we conclude this section by the 
presentation of the Hopf structure of $U^\lambda_\theta(\mathfrak{p})$, i.e. we give explicit 
formulas for counits, coproducts and antipodes for all solutions that are considered here and give 
the proof of the Hopf algebra axioms.

In parallel, as mentioned in the introduction, we sketch a first scheme of a general method that 
shall provide the opportunity to derive deformations $U_h(\mathfrak{g})$ to any given noncommutative 
spacetime algebra $\mathfrak{X}_h$ with deformation parameter $h$. Hence the line of our arguments 
is drawn in terms of a general Lie algebra $\mathfrak{g}$ and we treat 
$U^\lambda_\theta(\mathfrak{p})$ on $\mathfrak{X}_\theta$ as an example.

We emphasize that the development of this method is still a work in progress. Here we  merely want to draw 
the outline of our idea and show that already at this stage it can be applied successfully, as one of the
solutions that we present here, $U^{(\frac{1}{2}, 0)}_\theta(\mathfrak{p})$, was also achieved recently by 
alternative approaches \cite{wess}, \cite{chaichian}. Thus, many aspects that touch on to the presented 
scheme will be treated independently in our subsequent work. 

\subsection{Conditions for Deformations $U^\lambda_\theta(\mathfrak{p})$ as Representations over 
$\mathfrak{X}_\theta$}

Since deformations $U^\lambda_\theta(\mathfrak{p})$ of the Poincar\'e algebra $\mathfrak{p}$ are of 
quantum universal enveloping algebra type we first clarify such notions as that of $U_h(\mathfrak{g})$ and 
that of representations on a given spacetime algebra $\mathfrak{X}_h$ in terms of a 
general Lie algebra $\mathfrak{g}$.
\bdefinition 
        A p-dimensional \emph{Lie algebra} over the field $\mathbf{K}$ is a 
        $\mathbf{K}$-linear vector space endowed with a map
        $$\left[ \; , \; \right]: \mathfrak{g} \times \mathfrak{g} \longrightarrow \mathfrak{g}$$
        called \emph{bracket} with the following properties: 
        \beqarr
                \forall g, h, k \in \mathfrak{g}: 
                        & & \left[g, h\right] = - \left[h, g\right] \; \textrm{(Antisymmetry)} \nn \\
                        & & \left[g + h, k\right] = \left[g, k\right] 
                                         + \left[h, k\right] \; \textrm{(Linearity)} \nn \\
                        & & 0 = \left[\left[g, h\right], k\right] 
                                + \left[\left[h, k\right], g\right] 
                                + \left[\left[k, g\right], h\right]
                                \; \textrm{(Jacobi-Identity)} \nn
        \eeqarr
        Linearity holds for both components.
\edefinition
Since it is a vector space, the Lie algebra $\mathfrak{g}$ has a $p$-dimensional basis $(g)_{i\in I}$ with 
$I = \{1 \ldots p \}$. Hence the bracket $\left[ \; , \; \right]$ can be expressed as a linear 
combination of the basis elements in terms of the Lie algebra's structure constant 
$c_{ij}^k \in \mathbf{K}$. For all $i, j, k \in I$ we have then
$$\left[g_i, g_j\right] = i c_{ij}^k g_k.$$
Since direct sums and tensor products of vector spaces are vector spaces themselves, to any Lie 
algebra $\mathfrak{g}$ there exists the tensor or free algebra $\mathcal{T}(\mathfrak{g})$
$$\mathcal{T}(\mathfrak{g}) = \mathbf{K} \oplus \mathfrak{g} \oplus (\mathfrak{g}\otimes \mathfrak{g}) 
        \oplus \ldots \oplus \mathfrak{g}^{j \otimes} \oplus \ldots . $$
that leads us directly to the constructive definition of a universal enveloping algebra.
\bdefinition
        If $\mathfrak{g}$ is a Lie algebra with bracket $\left[ \; , \; \right]$ then the 
        \emph{universal enveloping algebra} $U(\mathfrak{g})$ of $\mathfrak{g}$ is the tensor algebra 
        $\mathcal{T}(\mathfrak{g})$ devided by the two-sided ideal $\mathcal{I}$
$$ U(\mathfrak{g}) = \frac{\mathcal{T}(\mathfrak{g})}{\mathcal{I}} $$
        that is generated by relations
$$ g_i \tensor g_j - g_j \tensor g_i - i c_{ij}^k g_k = 0.$$
\edefinition
The deformation of a Lie algebra $\mathfrak{g}$ is performed by quantizing its universal enveloping 
algebra that we denote by $U_h(\mathfrak{g})$. This is because the commutator bracket 
$\left[G_i \stackrel{\tensor}{,} G_j\right] = G_i \tensor G_j - G_j \tensor G_i$ for 
generators $(G_i)_{i \in I}$ of $U_h(\mathfrak{g})$ maps within $U_h(\mathfrak{g})$, i.e. the 
commutator in general is a linear combination in terms of the infinit dimensional basis 
of $U_h(\mathfrak{g})$ generated by $(G_i)_{i \in I}$. Thus $U_h ({\mathfrak{g}})$ becomes a Lie 
algebra with
\beqarr
        \left[ \; , \; \right]: U_h(\mathfrak{g}) \tensor U_h(\mathfrak{g}) 
                & \longrightarrow & U_h(\mathfrak{g}) \nn \\
        \left[G_i, G_j\right] & \mapsto & i C_{ij}(G_k,h). \label{Gcommutator}
\eeqarr
In the further consideration, we omit the symbol of tensor multiplication. The quantum universal 
enveloping algebra is thus defined to be the free associative algebra generated by $(G_i)_{i \in I}$ 
that is divided by the ideal $\mathcal{I}_h$ generated by (\ref{Gcommutator}) such that 
for $h \to 0: \; \mathcal{I}_h \to \mathcal{I}$ and consequently
$$ U_h(\mathfrak{g}) \to U(\mathfrak{g}). $$
We adjourn the discussion concerning the potential change of the number of generators $(G_i)_{i \in I}$
under deformations and exclude such solutions for $U_h(\mathfrak{g})$. 
To be well defined in this way the multiplication in $U_h(\mathfrak{g})$ has to satisfy 
closure and associativity. This is the first condition expressed by the Jacobi-Identity for 
the functions $C_{ij}(G_k,h)$
\begin{flushleft}
        \textsc{Condition 1}
\end{flushleft}
\beqarr
        0 & = & \left[\left[G_i, G_j\right], G_k\right] 
                + \left[\left[G_j,G_k\right],G_i\right] 
                + \left[\left[G_k,G_i\right],G_j\right] \nn \\
        & = & i \left[C_{ij}(G_l, h),G_k\right] 
                + i \left[C_{jk}(G_l, h),G_i\right] 
                + i \left[C_{ki}(G_l, h),G_j\right]. \nn
\eeqarr
We now apply Condition 1 to the example of $U_\theta(\mathfrak{p})$. The commutation relations 
of the Poincar\'e algebra $\mathfrak{p}$ are given by
\beqarr
   \left[p^\mu,p^\nu\right] &=& 0 \nn \\
   \left[m^{\mu\nu},p^{\rho}\right] &=& i\eta^{\mu\rho}p^\nu -i\eta^{\nu\rho}p^\mu \nn \\
   \left[m^{\mu\nu},m^{\rho\sigma}\right] &=&  i\eta^{\mu\rho}m^{\nu\sigma}- i\eta^{\nu\rho}m^{\mu\sigma}
                                            + i\eta^{\nu\sigma}m^{\mu\rho} - i\eta^{\mu\sigma}m^{\nu\rho}.
                                                                                                                  \label{poincare}
\eeqarr
For the case of canonical commutation relations the deformation is actually limited to the Lorentz 
algebra, such that the first relation of (\ref{poincare}) is preserved. As generators for 
$U_\theta(\mathfrak{p})$ we use momentum operators $P^\mu$ and Lorentz operators $M^{\mu\nu}$.
We make the following ansatz for the commutation relations of the deformed Poincar\'e algebra
\beqarr
        \left[P^\mu, P^\nu \right] & = & 0 \nn \\
        \left[M^{\mu\nu},P^\rho\right] & = & i (\eta ^{\mu\rho}P^\nu - \eta^{\nu\rho}P^\mu) 
        + i\chi_\theta^{\mu\nu\rho}(P^\gamma, M^{\kappa\lambda}) \nn \\
        \left[M^{\mu\nu}, M^{\rho\sigma}\right]
                & = & + i\eta^{\mu\rho}M^{\nu\sigma}- i\eta^{\nu\rho}M^{\mu\sigma}
                + i\eta^{\nu\sigma}M^{\mu\rho}- i\eta^{\mu\sigma}M^{\nu\rho} \nn \\
                & & + i\phi_\theta^{\mu\nu\rho\sigma}(P^\gamma,M^{\kappa\lambda}). \label{thetapoincare}
\eeqarr
The function $\phi_\theta^{\mu\nu\rho\sigma}(P^\gamma,M^{\kappa\lambda})$ is antisymmetric in the
first and second pair of indices and has physical dimension 1. The function 
$\chi_\theta^{\mu\nu\rho}(P^\gamma, M^{\kappa\lambda})$ is antisymmmetric in the first pair of 
indices and has physical dimension $\textrm{length}^{-1}$. Inserting the commutation relations 
(\ref{thetapoincare}) into the Jacobi-Identities corresponding to Condition 1
\beqarr
   0 & = & \left[\left[P^\mu, P^\nu \right], M^{\rho\sigma} \right]
        + \left[\left[P^\nu, M^{\rho\sigma} \right], P^\mu \right] 
        + \left[\left[ M^{\rho\sigma}, P^\mu \right], P^\nu \right] \nn \\
   0 & = & \left[\left[M^{\mu\nu}, M^{\rho\sigma} \right], P^\lambda \right]
        + \left[\left[M^{\rho\sigma}, P^\lambda\right], M^{\mu\nu}\right]
        + \left[\left[P^\lambda , M^{\mu\nu}\right], M^{\rho\sigma} \right]  \nn \\
   0 & = & \left[\left[M^{\mu\nu},M^{\rho\sigma}\right],M^{\kappa\lambda}\right]
        + \left[\left[M^{\rho\sigma},M^{\kappa\lambda}\right],M^{\mu\nu}\right] 
        + \left[\left[M^{\kappa\lambda},M^{\mu\nu}\right],M^{\rho\sigma}\right] \nn
\eeqarr
gives the following relations for the functions 
$\phi_\theta^{\mu\nu\rho\sigma}(P^\gamma,M^{\kappa\lambda})$  and 
$\chi_\theta^{\mu\nu\rho}(P^\gamma, M^{\kappa\lambda})$
\beqarr
   0 & = & [P^{\mu}, \chi_\theta^{\rho\sigma\nu}]-[P^{\nu}, \chi_\theta^{\rho\sigma\mu}] \nn \\
   0 & = & \left[P^\lambda, \phi_\theta^{\mu\nu\rho\sigma} \right]+
        [M^{\mu\nu}, \chi_\theta^{\rho\sigma\lambda}]-[M^{\rho\sigma}, \chi_\theta^
        {\mu\nu\lambda}] \nn \\
   0 & = & i\left[\phi_\theta^{\mu\nu\rho\sigma}, M^{\kappa\lambda} \right] 
        + i\left[\phi_\theta^{\rho\sigma\kappa\lambda}, M^{\mu\nu} \right] 
        + i\left[\phi_\theta^{\kappa\lambda\mu\nu}, M^{\rho\sigma} \right] \nn \\
   & & +\eta^{\nu\rho}\phi_\theta^{\mu\sigma\kappa\lambda}
        - \eta^{\mu\rho}\phi_\theta^{\nu\sigma\kappa\lambda} 
        + \eta^{\mu\sigma}\phi_\theta^{\nu\rho\kappa\lambda} 
        - \eta^{\nu\sigma}\phi_\theta^{\mu\rho\kappa\lambda} \nn \\
   & & - \eta^{\sigma\kappa}\phi_\theta^{\mu\nu\rho\lambda} 
        + \eta^{\rho\kappa}\phi_\theta^{\mu\nu\sigma\lambda} 
        - \eta^{\rho\lambda}\phi_\theta^{\mu\nu\sigma\kappa}
        + \eta^{\sigma\lambda}\phi_\theta^{\mu\nu\rho\kappa} \nn \\
   & & + \eta^{\lambda\mu}\phi_\theta^{\kappa\nu\rho\sigma} 
        - \eta^{\kappa\mu}\phi_\theta^{\lambda\nu\rho\sigma} 
        + \eta^{\kappa\nu}\phi_\theta^{\lambda\mu\rho\sigma} 
        - \eta^{\lambda\nu}\phi_\theta^{\kappa\mu\rho\sigma}. \label{mmp}
\eeqarr

Any pair of functions $\phi^{\mu\nu\rho\sigma}(P^\gamma,M^{\kappa\lambda})$ and 
$\chi_\theta^{\mu\nu\rho}(P^\gamma, M^{\kappa\lambda})$ solving these equations leads to 
well defined algebraic properties of $U_\theta(\mathfrak{p})$. 

Since $U_\theta(\mathfrak{p})$ shall be a representation on $\mathfrak{X}_\theta$, we now consider
the action of $G_i \in U_h(\mathfrak{g})$ on coordinates $x^\mu \in \mathfrak{X}_h$. To this purpose
the symmetry algebra has to be enhanced by a coalgebra structure.

\bdefinition 
        The \emph{coalgebra structure} on the $\mathbf{K}$-vector space $U_h(\mathfrak{g})$ 
        is given by the two linear operations counit $\counit$ and coproduct $\coproduct$. These are 
                  the maps
\beqarr
        \coproduct: U_h(\mathfrak{g}) & \to & U_h(\mathfrak{g}) \tensor U_h(\mathfrak{g}) \nn \\
        G_i & \mapsto & \coproduct (G_i) = \sum G_{i (1)} \tensor G_{i (2)} \nn \\
        \counit: U_h(\mathfrak{g}) & \to & K \nn \\
        G_i & \mapsto & \counit (G_i) \nn
\eeqarr
        that obey the two coalgebra axioms of counit and coassociativity
\beqarr
        (\counit \tensor \id) \circ \coproduct = & \id & = (\id \tensor \counit) \circ \coproduct \nn \\ 
        (\coproduct \tensor \id) \circ \coproduct & = &(\id \tensor \coproduct) \circ \coproduct. \nn
\eeqarr
\edefinition 

\bdefinition
        A \emph{bialgebra} is $\mathbf{K}$-vector space with algebra and coalgebra structure 
        made compatible by demanding that the coproduct and counit are algebra homomorphisms

\beqarr
        \coproduct(G_i G_j) = (\coproduct G_i)(\coproduct G_j), & \; & \coproduct \one = 
                  \one \tensor \one \nn \\
        \counit(G_i G_j) = \counit (G_i)\counit (G_j), & \; & \counit(\one) = 1. \nn
\eeqarr 
\edefinition
We now assume that $U_h(\mathfrak{g})$ is a bialgebra and consider its representations on 
$\mathfrak{X_h}$. Its associative multiplication for the coordinates
$x^\mu, x^\nu \in \mathfrak{X}_h$ is given by the commutator
$$ \left[x^\mu, x^\nu \right] = x^\mu x^\nu - x^\nu x^\mu = i \omega_h^{\mu\nu}(x^\rho). $$
Since the coordinates of $\mathfrak{X_h}$ are hermitean operators, the antisymmetric function
$\omega_h^{\mu\nu}(x^\mu)$ is real valued.
\bdefinition A \emph{representation} is a pair $(\rho, \mathfrak{X}_h)$ containing the homomorphism 
\beqarr
        \rho: \; U_h(\mathfrak{g}) & \to & gl(\mathfrak{X}_h) \nn \\
        G_i & \mapsto & \rho(G_i), \nn
\eeqarr
such that for $G_i, G_j \in U_h(\mathfrak{g})$ and $x^\mu, x^\nu \in \mathfrak{X}_h$
\beq
        \rho (G_i G_j - G_j G_i - i C_{ij}(G_k, h))x^\mu = 0 
        \label{algebrastructure} \\
\eeq
and
\beq
        \rho(G_i)( x^\mu x^\nu - x^\nu x^\mu - i \omega_h^{\mu\nu}(x^\rho)) = 0
        \label{spacestructure} \\
\eeq
are satisfied.
\edefinition
In other terms the algebraic structure of $U_h(\mathfrak{g})$ shall be represented in 
$gl(\mathfrak{X}_h)$ and these act as endomorphisms on $\mathfrak{X}_h$.

Introducing the left-action of $G_i \in U_h(\mathfrak{g})$ on coordinates $x^\mu \in \mathfrak{X}_h$ by
$$ G_i \laction x^\mu = \rho(G_i) x^\mu, $$
products of coordinates are mapped according to
\beqarr
        G_i \laction (x^\mu x^\nu) & = & \sum m \left(\left( G_{i(1)} \laction x^\mu \right) \tensor 
                \left( G_{i(2)} \laction x^\nu \right)\right) \nn \\
        G_i \laction \mathbf{1} & = & \counit(G_i)\mathbf{1}. \nn
\eeqarr
The multiplication $m$ is that of the coordinate algebra $\mathfrak{X}_h$. Before we continue
to actually define the action, we precede with some important remarks. In contrary to the 
commutative case, the commutator $[G_i, x^\mu]$ is not an element within $\mathfrak{X}_h$. We rather 
find that
$$ [G_i, x^\mu] \in \mathfrak{X}_h \tensor U_h(\mathfrak{g}). $$
Since $\rho(G_i) \in gl(\mathfrak{X}_h)$ the action cannot be defined by
the commutator $[G_i, x^\mu]$. This is only possible in the limit of $h \to 0$.

Thus we define the action of $G_i \in U_h(\mathfrak{g})$ 
on the coordinates $x^\mu, x^\nu \in \mathfrak{X}_h$ with $\mathbf{1} \in \mathfrak{X}_h$ by
$$ \rho\left(G_i\right) x^\mu = G_i \laction x^\mu := \left[G_i, x^\mu \right] \laction \one, $$
where the action of coordinates $x^\mu \in \mathfrak{X}_h$ on the unit element is merely a 
multiplication: $x^\mu \laction \mathbf{1} = x^\mu \one = x^\mu$. This way relation
(\ref{algebrastructure}) now reads
\beqarr
        \left(G_i G_j - G_j G_i - i C_{ij}(G_k, h) \right) \laction x^\mu & = & 0 \nn \\ 
        \Leftrightarrow
                \left(\left[G_i, \left[G_j, x^\mu \right] \right]               
 			- \left[G_j, \left[G_i, x^\mu \right]\right] -
                i \left[C_{ij}(G_k,h), x^\mu \right]\right) \laction \one & = & 0 \nn
\eeqarr
and thus we obtain
\begin{flushleft}
        \textsc{Condition 2}
\end{flushleft}
\beq
        \left[G_i, \left[G_j, x^\mu \right] \right]
                - \left[G_j, \left[G_i, x^\mu \right]\right]
                - i \left[C_{ij}(G_k,h), x^\mu \right] = 0. \nn
\eeq\\
Turning to relation (\ref{spacestructure}) we compute
\beqarr
	G_i \laction (x^\mu x^\nu - x^\nu x^\mu - i \omega_h^{\mu\nu}(x^\rho)) & = & 0 \nn \\
	\Leftrightarrow
	(\left[\left[G_i, x^\mu \right], x^\nu \right] 
                - \left[\left[G_i, x^\nu \right], x^\mu \right] 
                - i \left[G_i, \omega_h^{\mu\nu}(x^\rho) \right]) \laction \one & = & 0 \label{phi}
\eeqarr
and obtain 
\begin{flushleft}
        \textsc{Condition 3}
\end{flushleft}
\beq
        \left[\left[G_i, x^\mu \right], x^\nu \right] 
                + \left[\left[x^\nu, G_i \right], x^\mu \right] 
                + i \left[\omega_h^{\mu\nu}(x^\rho), G_i \right] = 0. \nn          
\eeq\\
To read off the bialgebra structure of $U_h(\mathfrak{g})$ we assume that the coproduct is 
of the general form
\beq
        \coproduct (G_i) = G_i \tensor \one + \one \tensor G_i + \sum \xi_{i(1)} \tensor \xi_{i(2)}. \nn
\eeq
We apply this again to the relation (\ref{spacestructure})
\beqarr
        0 & = & (G_i \laction x^\mu)x^\nu + x^\mu(G_i \laction x^\nu)
                        + \sum (\xi_{i(1)} \laction x^\mu)(\xi_{i(2)} \laction x^\nu) \nn \\
                & &     - (G_i \laction x^\nu)x^\mu - x^\nu(G_i \laction x^\mu)
                        - \sum (\xi_{i(1)} \laction x^\nu)(\xi_{i(2)} \laction x^\mu) \nn \\
                & &     - i G_i \laction \omega_h^{\mu\nu}(x^\rho)
\eeqarr
and compare with the computation (\ref{phi}) from above. We obtain for the coproduct
\beqarr
        \sum (\xi_{i(1)} \laction x^\mu)(\xi_{i(2)} \laction x^\nu) 
        & = & \left(\left[G_i, x^\mu \right] x^\nu \right) \laction \one
        - \left(\left[G_i, x^\mu \right] \laction \one \right)x^\nu 
        \label{condcoproduct} \\
        & = & \left(\left[\left[G_i, x^\mu \right], x^\nu\right] 
        + x^\nu \left[G_i, x^\mu \right] \right) \laction \one
        - \left(\left[G_i, x^\mu \right] \laction \one \right)x^\nu. \nn
\eeqarr
This formula will be used in the next subsection to compute the coproduct for the deformed Lorentz 
generators $M^{\mu\nu}$. For instance we turn again to the example $U_\theta(\mathfrak{p})$.
We make an ansatz for the commutator of $M^{\mu\nu}$ and coordinates 
$x^\rho \in \mathfrak{X}_\theta$. The corresponding relation for $P^\mu$ is of the classical 
form such that we have
\beqarr
        \left[P^\mu, x^\rho\right] & = & - i \eta^{\mu\rho} \nn \\ 
        \left[M^{\mu\nu}, x^\rho\right] & = & i(x^\nu\eta^{\rho\mu} - x^\mu\eta^{\rho\nu}) 
        + i\psi_\theta^{\mu\nu\rho}(P^\gamma,M^{\kappa\lambda}).
\eeqarr
The function $\psi_\theta^{\mu\nu\rho}(P^\gamma,M^{\kappa\lambda})$ has the physical dimension of 
length and is antisymmetric in the first two indices. Inserting this ansatz into Condition 2
\beqarr
   0 & = & \left[\left[P^\mu, P^\nu \right], x^\lambda \right] 
        + \left[\left[P^\nu, x^\lambda \right], P^\mu \right] 
        + \left[\left[ x^\lambda, P^\mu \right], P^\nu \right] \nn \\
   0 & = & \left[\left[M^{\mu\nu}, P^\rho \right], x^\lambda \right] 
        + \left[\left[P^\rho, x^\lambda\right], M^{\mu\nu} \right]
        + \left[\left[x^\lambda , M^{\mu\nu}\right], P^\rho \right]  \nn \\
   0 & = & \left[\left[M^{\mu\nu}, M^{\rho\sigma} \right], x^\lambda \right] 
        + \left[\left[M^{\rho\sigma}, x^\lambda\right], M^{\mu\nu}\right]
        + \left[\left[x^\lambda , M^{\mu\nu}\right], M^{\rho\sigma} \right]  \nn
\eeqarr
and replacing the commutators $\left[P^\mu, P^\nu \right], \left[M^{\mu\nu}, P^\rho\right]$ and 
$\left[M^{\mu\nu}, M^{\rho\sigma}\right]$ by their right hand sides, we obtain
\beqarr
   0 & = & [\psi_\theta^{\mu\nu\lambda}, P^\rho] 
        - [\chi_\theta^{\mu\nu\rho}, x^\lambda] \nn \\
   0 & = & i [M^{\rho\sigma}, \psi_\theta^{\mu\nu\lambda}] 
        - i [M^{\mu\nu}, \psi_\theta^{\rho\sigma\lambda}] 
        + i [\phi_\theta^{\mu\nu\rho\sigma}, x^\lambda] \nn \\
   & &- \eta^{\mu\rho}\psi_\theta^{\nu\sigma\lambda} + \eta^{\nu\rho}\psi_\theta^{\mu\sigma\lambda}
   - \eta^{\nu\sigma}\psi_\theta^{\mu\rho\lambda} + \eta^{\mu\sigma}\psi_\theta^{\nu\rho\lambda} \nn \\
   & &- \eta^{\sigma\lambda}\psi_\theta^{\mu\nu\rho} + \eta^{\rho\lambda}\psi_\theta^{\mu\nu\sigma}
   - \eta^{\mu\lambda}\psi_\theta^{\rho\sigma\nu} 
        + \eta^{\nu\lambda}\psi_\theta^{\rho\sigma\mu}. \label{mpx}
\eeqarr
Turning finally to Condition 3
\beqarr
   0 & = & \left[\left[P^\lambda , x^\mu \right], x^\nu \right]
        + \left[\left[x^\mu, x^\nu \right], P^\lambda\right]
        + \left[\left[x^\nu, P^\lambda\right], x^\mu \right] \nn \\
   0 & = & \left[\left[M^{\rho\sigma}, x^\mu \right], x^\nu \right]
        + \left[\left[x^\mu, x^\nu\right], M^{\rho\sigma}\right] 
        + \left[\left[x^\nu, M^{\rho\sigma}\right], x^\mu \right] \nn
\eeqarr
and replacing again by the corresponding right hand sides we obtain the single equation
\beq
   0 = i \left[\psi_\theta^{\mu\nu\rho}, x^\sigma \right] 
        - i \left[\psi_\theta^{\mu\nu\sigma}, x^\rho \right]
   - \eta^{\mu\rho}\theta^{\nu\sigma} + \eta^{\nu\rho}\theta^{\mu\sigma} 
   + \eta^{\mu\sigma}\theta^{\nu\rho} - \eta^{\nu\sigma}\theta^{\mu\rho}. \label{mxx}
\eeq
This final relation shows that the ansatz for $\psi^{\mu\nu\rho}$ can never be chosen to be 
zero or constant and such the coproduct of $M^{\mu\nu}$ is necessarily deformed. In the classical limit
$\theta^{\mu\nu} \to 0$ we have $\phi_\theta^{\mu\nu\rho\sigma}(P^\gamma, M^{\kappa\lambda}) \to 0$,  
$\psi_\theta^{\mu\nu\rho}(P^\gamma, M^{\kappa\lambda}) \to 0$ and 
$\chi_\theta^{\mu\nu\rho}(P^\gamma, M^{\kappa\lambda}) \to 0$.

Now we have obtained all conditions for $U_\theta^\lambda(\mathfrak{p})$ as a 
representation on $\mathfrak{X}_\theta$. In the next subsection we find solutions 
$U_\theta^\lambda(\mathfrak{p})$ by making an appropriate ansatz for the deformed Lorentz 
generator $M^{\mu\nu}$. 

\subsection{The Computation of Explicit Solutions $U^{(\lambda_1, \lambda_2)}_\theta(\mathfrak{p})$}

In general the functions $\phi_\theta^{\mu\nu\rho\sigma}(P^\gamma,M^{\kappa\lambda})$, 
$\psi_\theta^{\mu\nu\rho}(P^\gamma,M^{\kappa\lambda})$ and 
$\chi_\theta^{\mu\nu\rho}(P^\gamma, M^{\kappa\lambda})$ can be considered as power series 
in $\theta^{\mu\nu}$, given by
\beqarr
   \phi_\theta^{\mu\nu\rho\sigma}(P^\gamma,M^{\kappa\lambda}) 
        & = & \sum_{k=1}^{\infty} \phi_{\theta, k}^{\mu\nu\rho\sigma}(P^\gamma,M^{\kappa\lambda}) \nn \\
   \psi_\theta^{\mu\nu\rho}(P^\gamma,M^{\kappa\lambda}) 
        & = & \sum_{k=1}^{\infty} \psi_{\theta, k}^{\mu\nu\rho}(P^\gamma,M^{\kappa\lambda}) \nn \\
   \chi_\theta^{\mu\nu\rho}(P^\gamma,M^{\kappa\lambda}) 
        & = & \sum_{k=1}^{\infty} \chi_{\theta, k}^{\mu\nu\rho}(P^\gamma,M^{\kappa\lambda}). \nn
\eeqarr
The index of summation $k$ denotes the power of $\theta^{\mu\nu}$ in 
$\phi_{\theta, k}^{\mu\nu\rho\sigma}(P^\gamma,M^{\kappa\lambda})$, 
$\psi_{\theta, k}^{\mu\nu\rho}(P^\gamma,M^{\kappa\lambda})$ and 
$\chi_{\theta, k}^{\mu\nu\rho}(P^\gamma,M^{\kappa\lambda})$ 
respectively. We merely want to consider the most simple solutions to the set of equations (\ref{mmp}), 
(\ref{mpx}) and (\ref{mxx}) and thus we restrict ourselves to the case that is linear in $\theta^{\mu\nu}$. 
If we account for the physical unities, we find that
\beqarr
   \phi_\theta(P^\gamma,M^{\mu\nu}) = \phi_\theta(P^\gamma) & \sim & \theta P P \nn \\
   \psi_\theta(P^\gamma,M^{\mu\nu}) = \psi_\theta(P^\gamma) & \sim & \theta P \nn \\
   \chi_\theta(P^\gamma,M^{\mu\nu}) = \chi_\theta(P^\gamma) & \sim & \theta PPP. \nn
\eeqarr
Inserting this ansatz into the three conditions from the previous section  
generates the set of solutions in first order in $\theta$. An alternative method 
that gives the same resuts is assuming the deformed Lorentz generator $M^{\mu\nu}$ 
to be of the following general form
\beq
        M^{\mu\nu} = x^\mu P^\nu - x^\nu P^\mu + \Lambda^{\mu\nu}. \label{generator}
\eeq  
The function $\Lambda^{\mu\nu}$ has physical dimension 1 and is antisymmetric in its indices. 
It turns out in the next steps that any choice of $\Lambda^{\mu\nu}$ with these properties 
generates a valid solution of $U_\theta^\lambda(\mathfrak{p})$. We now express the functions 
$\psi^{\mu\nu\rho}_\theta(P^\gamma)$, $\phi^{\mu\nu\rho\sigma}_\theta(P^\gamma)$ and 
$\chi^{\mu\nu\rho}_\theta(P^\gamma)$  in terms of $\Lambda^{\mu\nu}$
\beqarr
        \chi_\theta^{\mu\nu\rho} & = & -i[\Lambda^{\mu\nu}, P^\rho] \nn \\
        \psi^{\mu\nu\rho}_\theta & = & \theta^{\mu\rho} P^\nu - \theta^{\nu\rho} P^\mu - 
                i\left[\Lambda^{\mu\nu}, x^\rho \right] \nn \\
        \phi^{\mu\nu\rho\sigma}_\theta & = & 
                - \eta^{\mu\sigma} \Lambda^{\rho\nu} - \eta^{\nu\sigma} \Lambda^{\mu\rho} 
                + \eta^{\mu\rho} \Lambda^{\sigma\nu} + \eta^{\nu\rho} \Lambda^{\mu\sigma}\nn \\
                & & + \theta^{\mu\rho} P^\nu P^\sigma - \theta^{\nu\rho} P^\mu P^\sigma 
                    - \theta^{\mu\sigma} P^\nu P^\rho + \theta^{\nu\sigma} P^\mu P^\rho\nn \\
                & & - i \left[\Lambda^{\mu\nu}, x^\rho \right] P^\sigma 
                + i\left[\Lambda^{\mu\nu}, x^\sigma \right] P^\rho 
                - i\left[x^\mu, \Lambda^{\rho\sigma}\right] P^\nu
                + i\left[x^\nu, \Lambda^{\rho\sigma}\right] P^\mu \nn \\
                & & + i x^\mu[\Lambda^{\rho\sigma},P^\nu]-ix^\nu[\Lambda^{\rho\sigma},P^\mu] 
                + i x^\sigma[\Lambda^{\mu\nu},P^\rho]-ix^\rho[\Lambda^{\mu\nu},P^\sigma] \nn \\
                & & - i \left[\Lambda^{\mu\nu}, \Lambda^{\rho\sigma}\right].
\eeqarr
Inserting these expressions into the conditions (\ref{mmp}), (\ref{mpx}) and (\ref{mxx}) results in
\beqarr
        0 & = & \left[ \left[\Lambda^{\mu\nu},\Lambda^{\rho\sigma} \right], P^\lambda \right] 
                + \left[ \left[\Lambda^{\rho\sigma}, P^\lambda \right], \Lambda^{\mu\nu} \right] 
                + \left[ \left[P^\lambda ,\Lambda^{\mu\nu} \right], \Lambda^{\rho\sigma}\right] \nn \\
        0 & = & \left[ \left[\Lambda^{\mu\nu},\Lambda^{\rho\sigma} \right], \Lambda^{\kappa\lambda} \right]
                + \left[ \left[\Lambda^{\rho\sigma},\Lambda^{\kappa\lambda} \right], \Lambda^{\mu\nu} \right]
                + \left[ \left[\Lambda^{\kappa\lambda},\Lambda^{\mu\nu} \right], \Lambda^{\rho\sigma} \right] \nn \\
        & & \nn \\
        0 & = & \left[ \left[\Lambda^{\mu\nu},x^\sigma \right], P^\rho \right] \nn \\
        & = & \left[ \left[\Lambda^{\mu\nu},x^\sigma \right], P^\rho \right] 
                + \left[ \left[x^\sigma, P^\rho \right], \Lambda^{\mu\nu}\right] 
                + \left[ \left[P^\rho, \Lambda^{\mu\nu}\right], x^\sigma \right] \nn \\
        0 & = & \left[ \left[\Lambda^{\mu\nu},\Lambda^{\rho\sigma} \right], x^\lambda \right] 
                + \left[ \left[\Lambda^{\rho\sigma}, x^\lambda \right], \Lambda^{\mu\nu} \right] 
                + \left[ \left[x^\lambda ,\Lambda^{\mu\nu} \right], \Lambda^{\rho\sigma}\right] \nn \\
        & & \nn \\
        0 & = & \left[\left[\Lambda^{\mu\nu},x^\rho \right],x^\sigma \right] 
                + \left[\left[x^\sigma,\Lambda^{\mu\nu} \right],x^\rho \right] \nn \\
        & = & \left[\left[\Lambda^{\mu\nu},x^\rho \right],x^\sigma \right] 
                + \left[\left[x^\sigma,\Lambda^{\mu\nu} \right],x^\rho \right]
                + \left[\left[x^\rho,x^\sigma \right],\Lambda^{\mu\nu} \right].
\eeqarr
Due to its physical dimension, the most simple structure of $\Lambda^{\mu\nu}$ is of the form
$$ \Lambda^{\mu\nu} \sim \theta P P. $$
Obviously any choice of $\Lambda^{\mu\nu}$ of this kind leads to a solution for 
$U_\theta^\lambda(\mathfrak{p})$. Omitting a possible constant, we choose $\Lambda^{\mu\nu}$ to be
\beq
        \Lambda^{\mu\nu} := \lambda_1 P_\alpha \left(\theta^{\mu\alpha} P^\nu - \theta^{\nu\alpha} P^\mu \right)
                + \lambda_2 \eta_{\alpha \beta} P^\alpha P^\beta \theta^{\mu\nu},
\eeq
with real parameters $\lambda_1, \lambda_2$. Computing now the functions $\psi^{\mu\nu\rho}(P^\gamma)$, 
$\phi^{\mu\nu\rho\sigma}(P^\gamma)$ \\ 
and $\chi^{\mu\nu\rho\sigma}(P^\gamma)$ with this expression, we obtain
\beqarr
        \chi^{\mu\nu\rho}_\theta & = & 0 \nn \\ 
        \psi^{\mu\nu\rho}_\theta & = & \left(1 - \lambda_1 \right)\left(\theta^{\mu\rho} P^\nu 
        - \theta^{\nu\rho} P^\mu \right)
        + \lambda_1 P_\alpha \left(\eta^{\mu\rho} \theta^{\nu\alpha} - \eta^{\nu\rho} \theta^{\mu\alpha} \right) 
        - 2 \lambda_2 \theta^{\mu\nu} P^\rho \nn \\
        \phi^{\mu\nu\rho\sigma}_\theta & = & 
        \left(1 - 2\lambda_1 \right)\left( \theta^{\mu\rho} P^\nu P^\sigma - \theta^{\nu\rho} P^\mu P^\sigma 
        - \theta^{\mu\sigma} P^\nu P^\rho + \theta^{\nu\sigma} P^\mu P^\rho \right) \nn \\
        & & - \lambda_2 \left( \theta^{\mu\rho}\eta^{\nu\sigma} - \theta^{\nu\rho}\eta^{\mu\sigma} 
        - \theta^{\mu\sigma}\eta^{\nu\rho} 
        + \theta^{\nu\sigma}\eta^{\mu\rho} \right)\eta_{\alpha\beta} P^\alpha P^\beta
\eeqarr 
and by this we have finally found all solutions $U_\theta^{(\lambda_1, \lambda_2)}(\mathfrak{p})$. 
The solution $U_\theta^{(\frac{1}{2}, 0)}(\mathfrak{p})$ gives the classical relations for 
the Lorentz algebra but with deformed coproduct - as we shall see in the next section. 
This result was also obtained by alternative considerations \cite{chaichian}, \cite{wess}. 

\subsection{The Hopf Algebra Structure of $U_\theta^{(\lambda_1, \lambda_2)}(\mathfrak{p})$}

In this final part we discuss the Hopf algebra structure of $U_\theta^{(\lambda_1, \lambda_2)}(\mathfrak{p})$.
Since the subalgebra of momentum operators $P^\mu$ is undeformed
$$ \coproduct(P^\mu) = P^\mu \tensor \one + \one \tensor P^\mu, 
\; \counit(P^\mu) = 0, \; S(P^\mu) = - P^\mu, $$
we merely focus on the corresponding properties for the deformed Lorentz generators 
$M^{\mu\nu}$. The necessary computations to prove the Hopf algebra axioms for $M^{\mu\nu}$ are 
straight forward, such that we limit ourselves to present the results and step through the necessary 
points without going into computational details. To ensure that counit and coproduct are algebra 
homomorphisms, they have to map the unit operator 
$\one \in U_\theta^{(\lambda_1, \lambda_2)}(\mathfrak{p})$ according to
\beqarr
        \counit(\one) & = & 1 \nn \\ 
        \coproduct(\one) & = & \one \tensor \one. \nn
\eeqarr
From relation (\ref{condcoproduct}) we read of the coproduct of $M^{\mu\nu}$ to be
\beqarr
        \coproduct(M^{\mu\nu}) & = & M^{\mu\nu} \tensor \one + \one \tensor M^{\mu\nu}
                - (1 - \lambda_1) P_\alpha \tensor \left(\theta^{\mu\alpha} P^\nu 
                - \theta^{\nu\alpha} P^\mu\right) \nn \\
                & & + \lambda_1 \left(\theta^{\mu\alpha} P^\nu - \theta^{\nu\alpha} P^\mu \right) \tensor P_\alpha
                + 2 \lambda_2 \theta^{\mu\nu}\eta^{\alpha\beta} P_\alpha \tensor P_\beta.
\eeqarr
Choosing the counit of $M^{\mu\nu}$ by 
$$ \counit (M^{\mu\nu}) = 0, $$
it is easy to see that $U_\theta^{(\lambda_1,\lambda_2)}(\mathfrak{p})$ satisfies the axioms of 
a bialgebra by proving the coalgebra axioms presented in the first subsection, such as counit 
and coassociativity
\beqarr
        (\counit \tensor \id) \circ \coproduct (M^{\mu\nu}) =
   & \id & = (\id \tensor \counit) \circ \coproduct (M^{\mu\nu}) \nn \\
        (\coproduct \tensor \id) \circ \coproduct (M^{\mu\nu}) 
   & = & (\id \tensor \coproduct) \circ \coproduct (M^{\mu\nu}). \nn
\eeqarr
To finally make $U_\theta^{(\lambda_1,\lambda_2)}(\mathfrak{p})$ a Hopf algebra, we need an antipode
$\antipode$ for $M^{\mu\nu}$. We find that
\beq
        \antipode(M^{\mu\nu}) = - M^{\mu\nu} - (1 - 2\lambda_1) (\theta^{\mu\alpha} P^\nu 
        - \theta^{\nu\alpha} P^\mu )P_\alpha 
        + 2  \lambda_2 \theta^{\mu\nu}\eta_{\alpha\beta} P^\alpha P^\beta
\eeq
satisfies the axiom for the antipode
\beq
        m \circ (\id \tensor \antipode) \circ \coproduct (M^{\mu\nu}) = \counit(M^{\mu\nu})\one
        = m \circ (\antipode \tensor \id) \circ \coproduct (M^{\mu\nu}), \nn
\eeq
where $m$ represents the multiplication within $U_\theta^{(\lambda_1,\lambda_2)}(\mathfrak{p})$. 
We remark that the double application of the antipode map $\antipode$ on $M^{\mu\nu}$ is the identity
operator
\beq
        \antipode^2 = \id. \nn
\eeq
The coproduct $\coproduct (M^{\mu\nu})$ and the antipode $S(M^{\mu\nu})$ for $\theta^{\mu\nu} \to 0$ 
converge to the undeformed case
\beq
   \coproduct (m^{\mu\nu}) = m^{\mu\nu}\tensor 1 + 1 \tensor m^{\mu\nu}, \; S(m^{\mu\nu}) = -m^{\mu\nu}, \nn 
\eeq
with $m^{\mu\nu} \in U(\mathfrak{p})$. Finally we have to ensure that coproduct and counit are 
algebra homomorphisms. Since the counit is trivial, the task reduces itself to satisfy the relations
\beqarr
   \left[\coproduct (M^{\mu\nu}), \coproduct (P^{\rho})\right]
        &=& i\eta^{\mu\rho}\coproduct (P^\mu) - i\eta^{\nu\rho}\coproduct (P^\nu) \nn \\
   \left[\coproduct (M^{\mu\nu}), \coproduct (M^{\rho\sigma})\right]
        &=& \;i \eta^{\mu\rho}\coproduct (M^{\nu\sigma}) - i \eta^{\nu\rho}\coproduct (M^{\mu\sigma}) 
                 + i \eta^{\nu\sigma}\coproduct (M^{\mu\rho})- i \eta^{\mu\sigma}\coproduct (M^{\nu\rho})  \nn \\
        & &      + i \coproduct (\phi^{\mu\nu\rho\sigma}) \nn 
\eeqarr
An easy computation shows that this is the case for all solutions 
$U_\theta^{(\lambda_1,\lambda_2)}(\mathfrak{p})$ that we have presented here.

\vspace{2em}

\section{Casimir Operators and Space Invariants}

In order to study field theoretic properties of the presented deformations $U^{(\lambda_1, \lambda_2)}(\mathfrak{p})$ 
we consider its central elements and spacetime invariants in this final section. 

Since the algebra of momenta $P^\mu$ is undeformed, the d'Alembert operator 
$\Box =\partial_\mu \partial^\mu=-P_\nu P^\nu$ and thus the Klein-Gordon operator are those of the classical case.

Concerning the Pauli-Lubanski vector and the spacetime invariant the situation is changed. 

We present a deformed Pauli-Lubanski vector $W^\lambda$ that transforms as a classical vector under the action of the Lorentz 
operators $M^{\mu\nu}$, such that its square is invariant under these operations. In order to obtain a spacetime invariant 
we find that the parameters $\lambda_1$ and $\lambda_2$ become dependent. 

\subsection{Pauli-Lubanski vector}

Since the commutation relations of the Lorentz generators $[M^{\mu\nu}, M^{\alpha\beta}]$ are deformed in general, 
the classical Pauli-Lubanski vector $\epsilon^{\lambda\kappa\rho\sigma}P_\kappa M_{\rho\sigma}$ does not
transform as a classical vector anymore
\beqarr
        [M^{\mu\nu}, \epsilon^{\lambda\kappa\rho\sigma}P_\kappa M_{\rho\sigma}]&=&i\eta^
        {\mu\lambda} \epsilon^{\nu \kappa\rho\sigma}P_\kappa M_{\rho\sigma}-i\eta^
        {\nu\lambda} \epsilon^{\mu\kappa\rho\sigma}P_\kappa M_{\rho\sigma}\nn \\
        && +2i\lambda_2
        (\theta^\mu_\alpha \epsilon^{\nu \alpha\lambda\rho}-\theta^\nu_\alpha \epsilon^
        {\mu \alpha\lambda\rho})P_\rho P_\beta P^\beta. \nn
\eeqarr
Moreover its square $W^\lambda W_\lambda$ turns out not to be invariant as well.
We define the deformed Pauli-Lubanski vector by the following properties
\beqarr
        \left[M^{\mu \nu}, W^{\lambda}\right]&= &i\eta^{\mu\lambda}W^\nu-i\eta^{\nu\lambda}W^\mu\nn\\ 
        \left[M^{\mu \nu}, W_{\lambda} W^{\lambda}\right] & =&  0 \nn\\ 
        \left[P^{\mu}, W_{\lambda} W^{\lambda}\right] & =&  0,
\eeqarr
and make an ansatz of the form $\epsilon^{\lambda\kappa\rho\sigma}P_\kappa 
M_{\rho\sigma}+(\epsilon\theta PPP)^\lambda$. We thus obtain the deformed Pauli-Lubanski vector to be
\beqarr
        W^\lambda=\epsilon^
        {\lambda\kappa\rho\sigma}P_\kappa M_{\rho\sigma}+\lambda_2 \epsilon^
        {\lambda\kappa\rho\sigma}\theta_{\kappa \rho} P_\sigma P_\alpha P^\alpha.
\eeqarr
It is remakable that $W^\lambda$ is independent of $\lambda_1$.

\subsection{Space Invariants}

Concerning the spacetime invariant $I$ we demand that it is merely an element of $\mathfrak{X}_\theta$. This is 
a strong requirement, since it becomes impossible to deform $I$ in any way. On the other hand the coproduct of 
$M^{\mu\nu}$ and thus its action on $I = x^\rho x_\rho$ is necessarily deformed, as we stated in reference to 
relation (\ref{mxx}).
We obtain for the action of $M^{\mu\nu}$ on $I = x^\rho x_\rho$ 
\beqarr
        M^{\mu\nu} \laction (x^\rho x_\rho) & = & ([M^{\mu\nu}, x^\rho x_\rho]) \laction \one\nn \\
        &=& ( -\theta^{\mu\nu} (2\lambda_2 n +4\lambda_1-2)-4i\lambda_2\theta^{\mu\nu}x^\rho P_\rho 
        -2i\lambda_1(\theta^{\mu\rho}x^\nu P_\rho - \theta^{\nu\rho}x^\mu P_\rho)\nn\\
        &&-2i(\lambda_1-1)(\theta^{\mu\rho}x_\rho P_\nu - \theta^{\nu\rho}x_\rho P_\mu) ) \laction \one \nn \\
        & = & -\theta^{\mu\nu} (2\lambda_2 n +4\lambda_1-2),
\eeqarr
where $n$ denotes the dimension of spacetime. To ensure that
$$ M^{\mu\nu} \laction I = 0 $$ 
we thus have to require that
\beq
        \lambda_1= \frac{1}{2}(1-n\lambda_2),
\eeq
and such the parameters $\lambda_1$ and $\lambda_2$ become dependent.

\vspace{2em}

\section{Conclusion}

In this paper we have derived a set of deformations  $U^{(\lambda_1,\lambda_2)}_\theta(\mathfrak{p})$ 
as representations on noncommutative spacetime algebras $\mathfrak{X}_\theta$. We have furthermore 
indicated how such deformations could possibly be constructed for arbitrarily given spacetime algebras.

At the moment it is still unclear whether the obtained solutions 
$U^{(\lambda_1,\lambda_2)}_\theta(\mathfrak{p})$ are related to each other. 
For physical reasons for example all solutions with $\lambda_2 = 0$ might turn out to be equivalent. 
Indeed, there are two hints in the present work that support this assumption. 
Firstly it is remarkable that the deformed Pauli-Lubanski vector is independent of $\lambda_1$, provided
that $\lambda_2 = 0$. With respect to particle states this means that the notion of spin does not 
depend on deformations of this kind.
Secondly, when the algebras $U^{(\lambda_1,0)}_\theta(\mathfrak{p})$ are represented on particle states
the function $\phi^{\mu\nu\rho\sigma}$ can be treated as a constant. 
So the contribution of this part of the deformation is likely to be only a phase factor similar to 
that generated by Moyal-Weyl star products, and one may speculate whether $\lambda_1$ could be treated 
as a $U(1)$ gauge parameter.

We would like to develop the ideas presented in this paper into a method that yields deformations
for any given noncommutative space. To this end it seems important to find a generating function for
the general case, such as $\Lambda^{\mu\nu}$ for the special case of 
$U^{(\lambda_1,\lambda_2)}_\theta(\mathfrak{p})$.

Finally, since the functions $\phi^{\mu\nu\rho\sigma}$ and $\psi^{\mu\nu\rho}$ depend on each other
it should be possible to derive the Hopf structure directly within the deformation procedure.

\vspace{2em}

\section{Acknowledgment}

We would like to express our gratitude to Julius Wess for his wise support that gave us the opportunity 
to complete this work.
We also thank Fabian Bachmeier and Jan-Mark Iniotakis for many fruitfull discussions and comments
as well as for carefully reading the manuscript.
This work was partially supported by the \emphbf{Ludwig-Maximilians-Universit\"at M\"unchen} within 
the program \emphbf{Gesetz zur F\"orderung des wissenschaftlichen und k\"unstlerischen Nachwuchses}.

\vspace{2em}


\newpage





\end{document}